# UBIQUITOUS HEALTHCARE MONITORING SYSTEM USING INTEGRATED TRIAXIAL ACCELEROMETER, $S_PO_2$ AND LOCATION SENSORS

Ogunduyile O.O[1]., Zuva K[2]., Randle O.A[3]., Zuva T[4]

[1,3,4] Department of Computer Science, Tshwane University of Technology, Pretoria, South Africa

[2] Department of Computer Science, University of Botswana, Gaborone, Botswana

## ABSTRACT

*Ubiquitous healthcare has become one of the prominent areas of research inorder to address the challenges encountered in healthcare environment. In contribution to this area, this study developed a system prototype that recommends diagonostic services based on physiological data collected in real time from a distant patient. The prototype uses WBAN body sensors to be worn by the individual and an android smart phone as a personal server. Physiological data is collected and uploaded to a Medical Health Server (MHS) via GPRS/internet to be analysed. Our implemented prototype monitors the activity, location and physiological data such as $SpO_2$ and Heart Rate (HR) of the elderly and patients in rehabilitation. The uploaded information can be accessed in real time by medical practitioners through a web application.*

## KEYWORDS

*Android smart phone, Ubiquitous Healthcare, Web application server, Wireless Body Area Networks*

## 1. INTRODUCTION

The challenges in healthcare system worldwide is to provide high quality service provisioning, easy accessibility and low cost service to ever increasing population particularly the elderly suffering from age related diseases. [1]; [2][3]. These challenges are placing a strain on the existing healthcare systems therefore they create the necessity to develop better, smarter, cost effective and healthcare systems to provide quality healthcare services at runtime. With this a large number of people, especially the elderly and those in rehabilitation will have easier access to the needed healthcare resources and quality-oriented healthcare services with limited financial resources. In recent times people are more health conscious and as noted by [4] the demand for better healthcare services is on the rise, individuals are demanding for better healthcare services that can be provided through ubiquitous healthcare systems.

Healthcare services that are availed to everyone independent of time and location is known as ubiquitous healthcare. Ubiquitous healthcare systems holds the potential of maintaining wellness, disease management, support for independent living, prevention and prompt treatment, along with emergency intervention anytime and anywhere as and when needed [5]. Moreover, technologies that provide ubiquitous healthcare services will be assimilated seamlessly in our daily lives such that they become invisible [6]. Ubiquitous healthcare systems use a large number of environments and platforms including Wireless Body Area Networks (WBANs), mobile devices and wireless grid/cloud/web services to make healthcare services available, observable, transparent, seamless, reliable and sustainable. Using these systems medical practitioners can remotely monitor, diagnose, access vital patient symptoms, offer advice to patients, facilitate real time communication with patients, give patients control over their personal data and also allow patients' access services anywhere anytime. Accessibility to several available services from an healthcare provider, flexibility, security and remote health data acquisitioning, service personalization, automatic decision making and response form the requirements for ubiquitous healthcare systems. WBANs are characterized by the deployment of biomedical sensors around human body to proactively collect the physiological data measurements and transmit them wirelessly to the base coordinator for processing [7]. WBAN environment can continuously interact with the neighbouring network nodes and can access services from the web/cloud/grid environment to provide new services at runtime. Providing an effective service provisioning is still a huge challenge in a WBAN environment.

The system prototype Service Oriented Wireless Body Area Networks (SOWBAN) developed in this study is an effective service provisioning mechanism in ubiquitous healthcare environment. It is designed to be used by patients in rehabilitation and to monitor the elderly, with the help of a 3 axis accelerometer and a wireless pulse oximeter ($SpO_2$). Body movements are monitored by an accelerometer to determine the patient's activities i.e., running, resting, walking and dangerous activities such as falling. The $SpO_2$ is used to measure blood-oxygen saturation levels ($SpO_2$) and Heart Rate (HR) [8]. The continuous observations of physiological data has the potential to greatly improve the quality of life of patients [9]. The hardware components of the prototype comprises of WBAN nodes (hereafter referred to as BS nodes or Body Sensor nodes), implemented on arduino fio platforms to collect physiological data and transmit it wirelessly to the Central Intelligent Node (CIN). From the CIN, physiological data of patient is then uploaded to the Medical Health Server (MHS). Physiological data received on the MHS would be equivalent to that obtained at the medical facility, if the patient were to go there for a medical check-up.

The prototype SOWBAN shows a system that implements fully the concept of ubiquitous healthcare service provisioning. The prototype architecture provides another solution to the telemedicine technology, which is high in cost in implementation and maintenance. Through smart designing methods, we have integrated certain services in our prototype and make them accessible at the users' ends (patients and medical practitioner). All events and processes within our prototype are designed as services. Such service oriented designs gives our prototype the advantages of: interoperability, reuse, efficiency and scalability. This current study aimed to describe our system architecture to integrate WBAN with WS for ubiquitous healthcare service provisioning. Our objectives are the following:

(a)   To enable the flawless incorporation of wireless body sensors with web service architectures.
(b)   To support the ubiquitous provisioning of high quality and cost-effective healthcare services in a distributed service provisioning environment.
(c)   To allow for real time diagnosis of the healthcare conditions of remote patients, independent of their location.

The remainder of this paper is succinctly summarized as follows. In Section 2, we discuss the related work. In Section 3, we discuss the prototype SOWBAN system architecture looking at its different layers to feature all hardware and software components. In Section 4, we present the live implementation of the prototype SOWBAN architecture. Section 5, concludes the paper and highlights its contributions as well as our future work.

## 2. RELATED WORK

Most of research has been put into remote vital signals acquisition using WBAN technologies. A survey of WBAN applications can be found in [10]. In particular, [11] developed the OnkoNet architecture to support healthcare services independent of time and place using mobile computing technology. A wearable health systems using WBAN for patient monitoring was introduced in three (3) layers that is; physiological sensors in first layer, personal server in second, and finally the health care servers and related services [12] . An ubiquitous Healthcare System (UHS) was developed by [13], the system consists of vital signs devices and environment sensor devices to acquire context information to monitor and manage health status of patients anytime anywhere. The framework targeted the development of four healthcare applications including self-diagnosis, remote monitoring, exercise management and emergency services. [14], proposed WBAN to support medical applications and the necessity for design concepts of the hardware and network protocols in a multi-patient monitoring environment were highlighted.

An information-based probabilistic relation model among the key indicators which sequenced their data gathering priority and precedence in the WBAN was constructed. They further constructed a cost function over the energy expenditure involved in their data gathering, and expressed the relationship between utility gain and energy loss as a constrained optimization problem. [15] gave a case study providing the fundamentals of how WBAN can be used for remote data acquisition and information fusion. Wireless devices from different technologies were made to work together in a distributed way in a smart environment. This system uses a distributed approach to add new components in execution time. [16], proposed a Secure Ubiquitous Healthcare System Architecture (SUHSA) to enable real time collection of healthcare data from WBAN. The collected data is converted into a Clinical Document Architecture (CDA) format, digitally signed, encrypted and securely transmitted over the Internet protocol Multimedia Subsystem (IMS) and Health Level 7 (HL7) messaging standards to a central hospital for patient health condition to be accessed by doctors. The IMS provides internet services with Quality of Service (QoS) and it integrates different services.

A comprehensive discussion of the roles of telemedicine, wireless body area networks and wireless utility grid computing technologies to address the challenges of the conventional healthcare system can be found in [17]. Their work integrated telemedicine technologies with Mobile Dynamic Virtual Communities (MDVC) into the healthgrid for cost-effective, quality and ubiquitous healthcare service provisioning. The framework allowed for remote vital signs acquisition and personalized grid services discovery through a metadata and fuzzy logic based intelligent context engine. The gFrame is promising because of its vision to integrate MDVC into the health grid for ubiquitous, quality and cost-effective healthcare service provisioning. This paper reports on life prototype implementation of SOWBAN, which is a component of gFrame for remote vital signs acquisition. The results of this study were obtained from volunteers.

## 3. SOWBAN SYSTEM ARCHITECTURE

The system architecture of SOWBAN consists of the following layers: BS Arduino Layer, Central Intelligent Node Layer (CINL), and Medical Health Server (MHS) Layer which is made up of the SOWBAN applications (the web application server) and the database as shown in Figure 1.

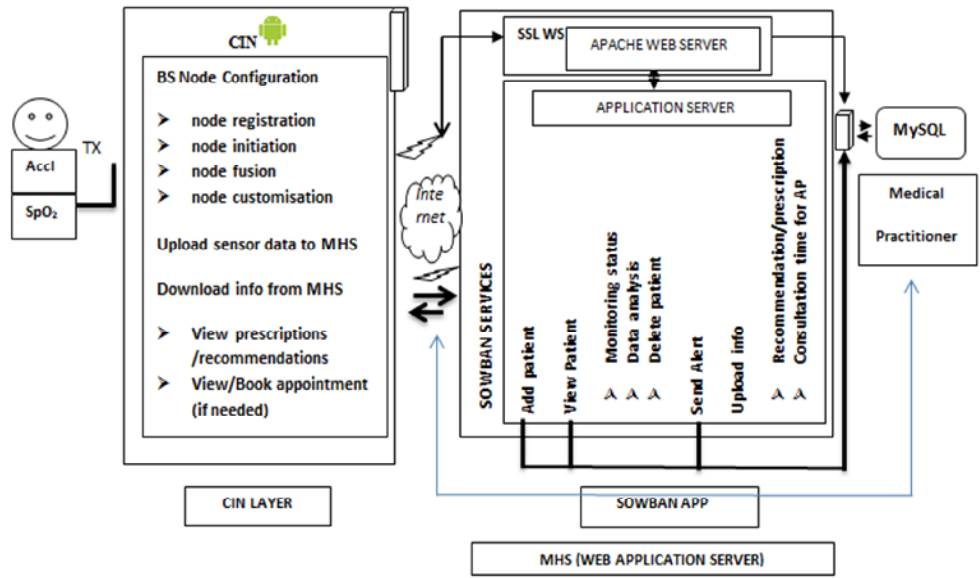

Figure 1. SOWBAN Architecture

### 3.1. The BS Arduino Layer

The BS Arduino layer consists of the BS nodes i.e., 3 axis accelerometer and pulse oximeter, implemented on arduino fio platforms to be worn on the patient's body. The BS nodes gather physiological data from patients and transmit the data through an attached WiFly radio over a Wi-Fi wireless network to the CINL. The BS nodes have the intelligence to sense, sample, process and communicate physiological data. They also satisfy design requirements of minimum weight, greatly reduced form factor, low – power consumption through configuration (which allows for prolonged ubiquitous healthcare monitoring), seamless integration into our prototype, standard based interface protocols and patient – specific customization. The BS nodes receive configuration instructions and responds to commands from the CIN which is more superior in terms of intelligence [18].

### 3.2. Central Intelligent Node Layer (CINL)

The CINL holds the CIN which is responsible for collecting and processing of physiological data received from BS nodes, providing a Graphical User Interface (GUI) for the patient to view his/her data in real time and uploading patient data to the MHS. It communicates with the MHS via GPRS/Internet. The architecture uses an android smart phone with operating system (OS) 2.3v as the CIN. Android OS serves as a great choice for this architecture allowing for sophisticated real time data processing and increasing processing power. Integrating an android smart phone with certain sensors e.g., an accelerometer provides a way to determine mobility and Global Positioning System (GPS) for location determination. This makes them more suitable for a fully integrated BS node ubiquitous monitoring system and is a benefit to our prototype SOWBAN. We have deployed a service on the CINL to enable configuration of the BS nodes, in terms of**:** node registration, node initialization, node fusion and node customisation. We have also deployed services to enable patients upload physiological data to the MHS, download information (recommendations/prescriptions) from the MHS and view the medical practitioner's consultation times or routine, if uploaded to the MHS. This will enable patients' book medical appointments when needed.

Using a simple comparison algorithm the CIN compares changes in data received from the BS nodes to determine whether to send data to the MHS or not. Data from the BS nodes are sent only when the CIN detects a change in the received data ($SpO_2$ parameters, activities and location). This will help save cost for the patient by reducing the number of data transmissions made. Figure 2 shows the basic data flow algorithm of the CINL.

```
Do
If (bsData 1 == ())
        bsData1 = readValue ()
        bsData2 = bsData1
Else
        bsData1 = readValue ()
        // compare with previous and decide
            If (bsData1! = bsData2)
            SendnewVal
            bsData2 = bsData1
            Endif
Endif
```

Figure 2. CINL basic data flow algorithm

### 3.3. Medical Health Server (MHS) Layer

The MHS stores the electronic medical records of registered users and provides various services to the patients, medical practitioners and healthcare providers. It is the responsibility of the MHS to authenticate users, accept health monitoring session uploads, format and insert the session data into corresponding patient medical records, analyse the data, determine serious health situations in order to contact emergency healthcare providers, and forward new information to the patients, such as recommendations, prescribed drugs and exercises. These operations are performed autonomously, without human's intervention by comparing the latest patient's physiological data updates with already existing ones in patient's medical record as well as comparing recommendations and prescriptions by the patient's medical practitioner or healthcare provider. The MHS consists of the web server and application server forming a web application server which can be accessed through a web browser. The application component of the web application server holds the MHS's business processes and is responsible for communications between the CIN, web server/web services and the database.

Our developed web application server is deployed on Tomcat with Apache and MySQL is used as a database. On the application component, Add patient, registers and initializes a new patient record on the database. View patient, checks the existing patient medical record i.e., monitoring status/monitoring history, received data analysis and if a patient record no longer exists on the database, delete patient. Send alert, creates an alert when data analysis results in a potential medical condition. With the Upload info, patient recommendations, prescriptions and medical practitioner's consultation slots can be uploaded to the MHS. Patients can view this uploaded information through the download info service on the CIN over the web.

Two main services offered at the web server component are *enterData* and *collectData*. To access these services the medical practitioners have to login first. Using the *enterData* enables data to be entered into the database while *collectData* helps retrieve data. These services give the medical practitioner access to the patient's record on the MHS over any public network provided the correct authentication details are used. Authentication login details ensure the privacy of the patient's medical records. Privacy involves the confidentiality of patient's data and the assurance that no information leakage from the medical records is feasible.

## 4. SOWBAN IMPLEMENTATION

The process of monitoring a person's health condition requires a large amount of information. Our developed BS nodes implemented on arduino fio platforms has the capacity to collect large amount of physiological data and transmits these data continuously. Figure 3 shows the diagrammatical description of the data transmission process of our prototype.

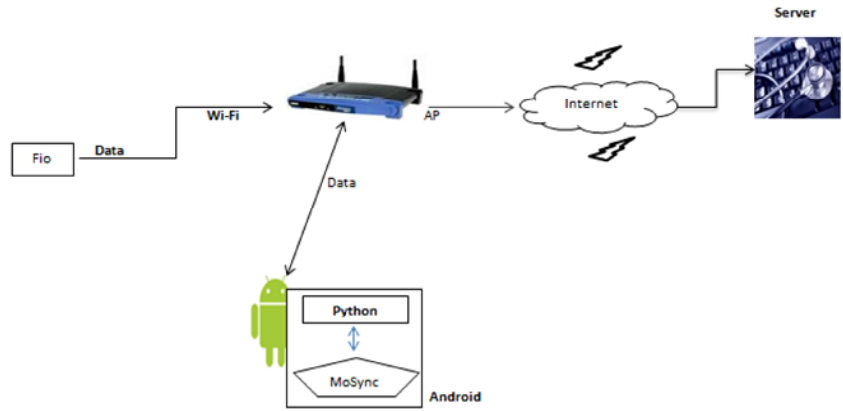

Figure 3. Prototype SOWBAN data transmission process

The accelerometer is worn around the patient's waist, either using a waist belt or modified mobile-phone carry-case. The accelerometer gives the continuous monitoring and transmission of the patient's angular force on the X, Y and Z axis. These movements are translated into motion states which corresponds to the activities of the patient with the following states existing; resting (ID - 1), walking/general movement (ID - 2), running (ID - 3) and falling (ID – 4). The measurement of the motion is determined by using the maximum recorded values from the accelerometer within a 1- second cycling period and based on these values, the respective states were determined. A workflow process was designed for the accelerometer, as shown in figure 4.

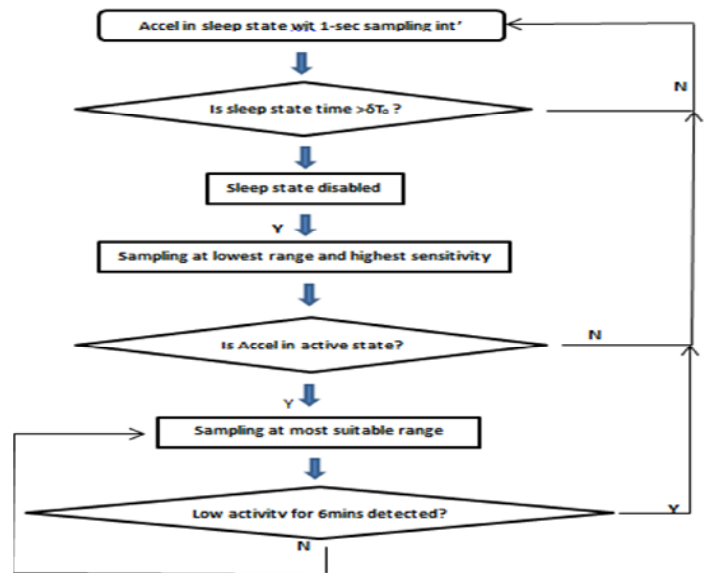

Figure 4. Workflow process of the accelerometer

Here the accelerometer is initially set in sleep state and its timer is configured to wake up at 1-sec intervals collecting samples at both low measurement range (i.e., ±2g) and high sensitivity on the X, Y and Z axis. This conserves power when person is in a resting state (ID – 1). The sleep state is disabled once a measurement value exceeds a certain threshold value ($T_o$) because this signifies person is now in an active state and the accelerometer continues collecting samples in this state. Since the normal frequency of human activities is at ranges between 1 – 18Hz as discovered by [19], the data sampling rate is set to be in the range of 10 – 100Hz.

Measurement range of the accelerometer is set to adapt to the exact accelerometer measurement. For example, if the measurement range is ±4 g full resolution, with sensitivity of 128LSB/g and the exact measurement is below -4g or above 4g, then the next measurement will be taken from a higher range of ±8 g with lower sensitivity of 64LSB/g and if exact measurement is between -2 g and 2g, then the next measurement will be taken with a lower range ±2 g with higher sensitivity of 256LSB/g. These measurement ranges allows for highest sensitivity outputs on the X, Y and Z axis while getting the closest accelerometer reading. Additionally, measurement ranges are adjusted independently in the different axis**.** The accelerometer's measurement ranges allows for highest sensitivity outputs on the X, Y and Z axis. If measurement shows inactivity for a period of time (e.g., 6mins) the accelerometer returns to an inactive state and restarts the work flow process all over. This work flow process ensures automatic activity monitoring, power conservation and alertness to changes in patient activities.

To carry out measurement test, our accelerometer, sampling at a frequency of 50Hz on all axes is worn around a person waist using a modified mobile-phone carry-case. Figure 5 shows the test measurements for the ID – 2 and ID – 3 states. We use AcclX, AcclY and AcclZ to represent the front, side and vertical accelerations, respectively. Due to the earth's gravitational force, there is always an output of +g present if a person is in a vertical position and vertical measurement AcclZ is taken. During test measurements, ID – 1 and ID – 2 showed acceleration movements patterns mostly on the X and Y axis while the Z axis showed less acceleration movement patterns. Also, measurement tests for ID – 3 and ID – 4 showed acceleration movements patterns mostly on the Z axis while the X and Y axis showed less acceleration movement patterns.

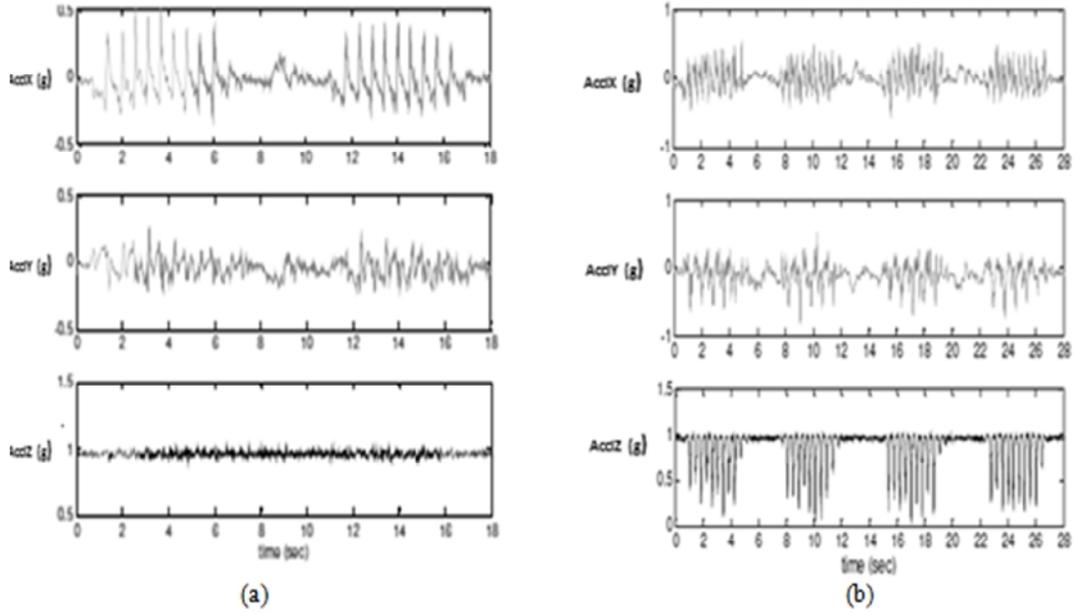

Figure 5. Accelerometer measurements showing (a) walking (ID – 2) and (b) running (ID – 3)

The measurement tests shows, different human activities generate different patterns in acceleration readings. With this, it is possible to group different human activity through the analysis of the recorded acceleration data and warn against abnormal activities such as increased energy expenditure and falling. Moreover, once an abnormal activity is detected through the acceleration reading, the health status can be verified with inputs from other sources, e.g., by checking the heart rate from a $SpO_2$ or heart activity from an ECG sensor.

The wireless pulse oximeter measures blood-oxygen saturation levels ($SpO_2$) as well as heart rate (HR). Measurements are based on Lambert Beer's law of spectral analysis which relates the concentration of absorbent in solution to amount of light transmitted through the solution [20]. Knowing the intensity, the path length and extinction co-efficient of a substance (here, oxyhemoglobin or reduced hemoglobin) at a particular wavelength, we determine oxygen saturation by measuring the light transmitted at two different wave lengths through the fingertip. This method capitalizes on the fact that, at the red region of the light spectrum around 660nm reduced hemoglobin (Hb) has higher extinction co-efficient compared to oxyhemoglobin ($HbO_2$). While in the near – infrared region of the light spectrum around 940nm, the extinction co-efficient of Hb is low compared to $HbO_2$. With the differences in extinction co-efficient, the light absorbed by Hb and $HbO_2$ can be determined and used to calculate a ratio R, which correlates to oxygen saturation. Using lambert's law, we obtain an expression for the ratio for light absorbed given by Equation 1.

$$R = \frac{\log_{10}(I_1/I_{in1})}{\log_{10}(I_2/I_{in2})} \quad (1)$$

$I_1$ and $I_2$ are the light intensity at different wave length calculated by Equation 2 and Equation 3

At wavelength $\lambda_1$, $\quad I_1 = I_{in1} 10^{-(\alpha_{o1}C_o + \alpha_{r1}C_r)l}$ (2)

At wavelength $\lambda_2$, $\quad I_2 = I_{in2} 10^{-(\alpha_{o2}C_o + \alpha_{r2}C_r)l}$ (3)

Where
  $C_o$ is the concentration of oxyhaemoglobin ($HbO_2$)
  $C_r$ is the concentration of reduced oxyhaemoglobin (Hb)

$\alpha_{on}$ is the extinction co-efficient of HbO2 at wavelength $\lambda_n$
$\alpha_{rn}$ is the extinction co-efficient of Hb at wavelength $\lambda_n$

From Equation 1 we show that:

$$\text{Oxygen saturation (SpO}_2\text{)} = \frac{C_o}{C_o + C_r} = \frac{\alpha_{r2} R - \alpha_{r1}}{(\alpha_{r2} - \alpha_{o2}) R - (\alpha_{r1} - \alpha_{o1})} \quad (4)$$

By measuring the elapsed time between peaks of the infrared light signal we get a value for heart rate using Equation 5. The infrared light signal is used to calculate for heart rate because it has low noise and can be used in different environments [21].

$$\text{Heart Rate (BPM)} = \frac{60}{Period(\sec)} \quad (5)$$

Our pulse oximeter test calculations report hearts rates in the range of 30 – 245bpm and $SpO_2$ values from 0 – 97%. In developing our wireless microcontroller based pulse oximeter, we made use of available off the shelf products that provide self-contained logics for driving LEDs (red, infrared, silicon photodiode) and performing $SpO_2$/HR calculations. The pulse oximeter performs all required calculations and transmits physiological data via a WiFly radio module over the Wi-Fi network to the CIN. The WiFly is attached to the arduino fio platforms to enable radio transmission. The WiFly can connect to any router whenever the necessary security protocols are correctly configured. Once connected, the module will create a server and listens on IP Address – 192.168.0.68 and Port – 60000.

The performance of the wireless network is important in pervasive healthcare service provisioning. We set up our Wi-Fi wireless network to ensure efficient transmission of patient physiological data because medical practitioners can only assess patient's status correctly by the medical information they receive. To achieve this, we configured an Access point (AP) to set up a wireless connection and deployed TDMA schemes to operate on the Wi-Fi's existing MAC protocols. Our TDMA schemes results in: a scheduled data transmission process, reduction in network congestion, power conservation and extended battery life of BS nodes.

We developed a MoSync python application to run on the android smart phone. Python for MoSync is a very effective programming language with efficient high-level data structures that delivers a simple but effective approach to object-oriented scripting programming [22]. The MoSync python application gathers data sent from the BS nodes. The CIN uses the python APIs (Application Program Interfaces) to manage the AP's processes i.e., setting up wireless connections to both BS nodes and the MHS**.** On receiving data from the BS nodes, the CIN uses comparison algorithms to compare data received with the last data sent. The comparison algorithm is programmed to send data to the MHS only when there is a difference in data just received from the last data sent. The CIN also enables the patient view his/her medical status in real time through its GUI. Figure 6 shows screenshots of the GUI interface on the CIN.

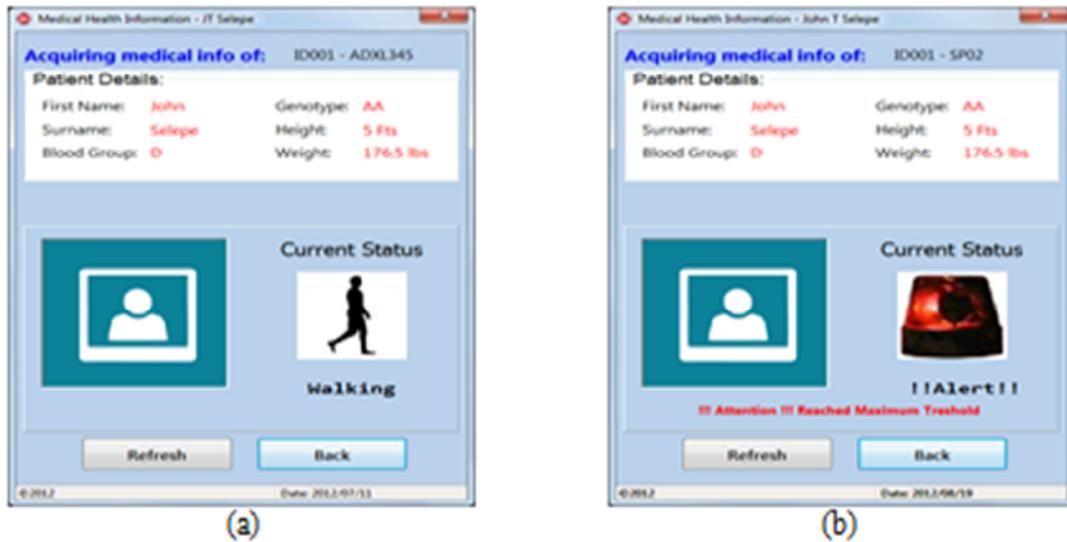

Figure 6. Screenshots of the CIN's GUI showing (a) patient walking (ID – 3) on accelerometer monitoring and (b) an alert message in case of an emergency on pulse oximeter monitoring

To upload patient data to the MHS, the CIN connects with the upload info service on its GUI and sends to the web application server in TCP/HTTP over GPRS/Internet. The web application server mines data received and determines health risks using logistic regression. Functionality on the web application server is implemented in HTML/PHP. The medical practitioners can see the patient's activities and $SpO_2$ health status in real time through a web application over a web browser. An abnormal condition triggers the alert signal which is received by both patient and medical practitioner and in high risk situations the GPRS on the android can be used to determine the patient's location. Our implementation of various off the shelf components and Open Source Softwares (Python, PHP, MySQL and Apache Tomcat) helps reduces overall cost of the prototype SOWBAN.

The power consumption and size for the major components in the BS Arduino Layer and CINL are listed in Table 1 to give a cost evaluation. The size of the entire BS arduino layer is limited to 90 *62* 22mm and powered by a Li-Ion battery pack which is selected due to its low cost and convenience in testing and recharging. With smart implementation power consumption is greatly reduced in the prototype development. The CINL has a more reliable and longer lasting source of power, therefore will not be an issue.

Table 1. Cost evaluation of the prototype SOWBAN

| Component | Current (mA) | Size (mm) |
|---|---|---|
| BS nodes i.e., accelerometer and pulse oximeter | 0.023 | 3 * 5 * 1<br>90 * 60.5 * 22 |
| Arduino Fio platform | 40 | 27.9 * 66.0 *3.5 |
| WiFly radio transmitter | 38 | 27*18*3.1 |
| Li-Ion battery pack | _ | 67.5*49.5*17 |

The cost to build the prototype SOWBAN (excluding the MHS) is around 1450 rands, which makes it feasible for ubiquitously monitoring the elderly and patients in rehabilitation. Based on our implementation, we can evaluate our prototype in terms of:

(a) Safety – patient's conditions being monitored constantly and in real time guarantees early detection in patient's status and treatment can be administered early enough.
(b) Timeliness – with real time detection in the patient's status i.e., especially in critical conditions where the medical practitioner is notified in immediately and can contact the healthcare providers in closest proximity to the patient for immediate attention.
(c) Effectiveness – by monitoring patient's status and providing services ubiquitously, cost spent on healthcare is reduced, patients do not have to travel the distance to a health facility, wait in queues at the health facility before being attended to. Effectiveness also ensures that patient's conditions, prescriptions and recommendations are properly documented.
(d) Efficiency – this is experienced in the ability for the designed prototype to provide an overall outcome for the patients in days of survival, years of survival, improvement in status (especially in recovering patients) and reduction in disability (especially in some age – related illness).

## 5. CONCLUSION

This paper describes the implementation of a Service Oriented Architecture (SOA) that seamlessly combines Wireless Body Area Network (WBAN) with Web services (WS) for ubiquitous healthcare service provisioning. The prototype SOWBAN developed proactively collects body signals of remote patients to recommend diagnostic services. This prototype provides continuous physiological data monitoring capabilities with minimum intervention of medical personnel and ubiquitous accessibility to variety of services allowing distributed healthcare resources to be massively reused for providing cost-effective services without individuals physically moving to the locations of those resources. By automating the physiological data monitoring process the most updated information of patient will be available at all times. The SOWBAN enables the monitoring of patients physiological data in real time, promising ubiquitous, yet an affordable and effective way to provide healthcare services. The developed SOWBAN is one of the first that successfully integrates body area networks with web services for ubiquitous healthcare service provisioning. The SOWBAN ensures interoperability between heterogeneous devices and technologies, data representations, scalability and reuse. The contributions of this paper are:

(a) It provides the effective provisioning of services within ubiquitous healthcare system. SOWBAN achieves this by seamlessly combining the WBAN technologies i.e., sensor systems and networks, embedded engineering etc. with web services i.e. internet technologies, apache servers etc.
(b) It allows health data monitoring, through activity information such as body positions and $SpO_2$ parameters, to know a patient's health status.
(c) It provides reliable and power efficient medical data transmission within the 802.11/Wi-Fi using developed TDMA schemes.

The SOWBAN also ensures privacy, integrity and authentication protocols by using passwords and encryptions because privacy is seen as a fundamental right of humans and is a sensitive issue in healthcare applications. However, future works are required to improve the quality of service of body sensors wireless communication, reliability of sensor nodes, security and standardization of interfaces. In addition, further studies of different medical conditions in clinical and ambulatory settings are necessary to determine specific limitations and possible new applications of this technology.

**Authors**

[1]Ogunduyile O.Oluwagbenga, is a lecturer at the Tshwane University of Technology, Pretoria, South Africa

[2]Zuva T, is a lecturer at the Tshwane University of Technology, Pretoria, South Africa

[3] Randle O. Abayomi, is a lecturer at the Tshwane University of Technology, Pretoria, South Africa.

[4]Zuva K, is a lecturer at the University of Botswana, Gaborone, Botswana